# I-INTERACTION: AN INTELLIGENT IN-VEHICLE USER INTERACTION MODEL

Li Liu[1] and Edward Dillon[2]

[1,2]Department of Computer Science, University of Alabama, Tuscaloosa, Alabama

[1]lliu@cs.ua.edu

[2]edillon@cs.ua.edu

## ABSTRACT

*The automobile is always a point of interest where new technology has been deployed. Because of this interest, human-vehicle interaction has been an appealing area for much research in recent years. The current in-vehicle design has been improved but still possesses some of the design from the traditional interaction style. In this paper, we propose a new user-oriented model for in-vehicle interaction model known as i-Interaction. The i-Interaction model provides user with an intuitive approach to interact with the In-Vehicle Information System (IVIS) by the keypad entry. It is the intent that the proposed usability testing for this model will help improve the way research and development is implemented from this topic. This model does not only provide the user with a direct interaction in vehicles but also introduce a new prospective that other research has not addressed.*

## 1. INTRODUCTION

Since Karl Benz's first "wheeled motor vehicle" for transporting passengers, humans continue to evolve one of our main sources of transportation with their intelligence. New engines and transmissions are invented along with the emergence of ecofuel are being employed. Due to the increase in the automobile to person ratio over the years, these motor vehicles have become more than a simple mean of transportation. Nowadays, people are performing tasks, other than driving, in their automobiles for different purposes such as navigating [1] and mobile communication through cellular devices [2]. Automobile manufacturers throughout time are designing and installing systems for both mechanical and electrical control to convenience drivers, like cruise control and climate control. At the same time, drivers expect to get more information while driving whether it is from the vehicle itself or the rest of the world. For information concerning the vehicle, since all control signals are transformed to the digital realm, [1] it is not hard for automobile designers to integrate new indicators and display screens on to both the central panel and instrument panel to inform the driver about the current condition of a vehicle, such as a tire pressure monitor, fuel economy calculator, etc. For the information outside the car, drivers are also offered a wide array of new equipment like GPS navigation system, cell phone merging system, and in-car entertainment systems. These sources of information no longer make a car to be some source of transportation but also a place for information access. Overall, it changes how we interact with the automobile.

Designing the interaction system for an automobile is challenging since a new prototype should be a fully functional proof-of-concept solution and should be rapidly developed due to the business opportunity of the automobile [3]. The new design also needs to be accepted by a great diversity of drivers. Researchers have discussed the In-Vehicle Information System (IVIS) [3, 4, 5, 6]. This system demonstrates a multimodal presentation system including the contents both shown on the central panel and the instrument panel. Some of the research concerning this system has been done in regards to the safety of drivers while interacting with an IVIS or just the general aesthetics of these systems. In regard to safety, there have been studies to measure the amount of attention used to concentrate on driving while interacting with some IVIS [4, 5, 6]. In relation to the aesthetics of an IVIS, there are studies that either have been done or currently being done to observe these systems. To be specific, there are IVIS that require voice interaction, [7, 8, 9] touch interaction [10], and even some that entail a certain hand gesture [11, 12, 13, 14].

Besides the information presented when a car is running, the IVIS can also show the condition of a car when a car is not running, for example, the vehicle service reminder system. The information that reveals the state of an automobile is shown by this system and is mostly used to determine if something on the vehicle should be serviced or not. This system is usually factory-programmed with a certain vehicle service interval. The system uses an internal clock or wheel speed sensor to record continuous data concerning distance. This data is calculated by an in-car CPU periodically. When the time or distance has exceeded the vehicle service intervals, the vehicle service reminder systems generates a vehicle service reminder indication on the display. Fortunately, some of today's vehicles already come with a built-in maintenance indicator that has related behaviors.

This kind of information from an IVIS is very important for both the well-being of the vehicle and cautionary purposes for the driver. Some systems employ information from the vehicle service reminder systems to improve driving safety. For example, in combination with the GPS navigation system, the car can direct the driver to an appropriate vehicle service location [15]. After a service is done to the vehicle, a reset operation needs to be done manually by an automobile technician. However, a technician may not be familiar with how to reset vehicle service reminder systems for different cars. Another situation is the fact that some people prefer to service their cars themselves. In both cases, people have to read the user manual to follow the step-by-step procedure to reset the service indicators.

At the same time, other computer controlled sub-systems which are also important to driving safety require input from the driver and technician, like the auto door lock and safe belt reminder at a certain speed, for example. Systems of this nature are typically used when the car is at rest. Those systems are able to give feedback to driver as output. In this paper, we study the IVIS in the case when the driver is not driving. We propose a new interaction model based on existing parts of the IVIS to solve the problem that these computer-controlled systems may solve but with a poor user interface.

A simulation of the i-Interaction shows that the new model is practical and will bring convenience to users because of the nature of a well-designed user interface. The usability test will be done as the future work while the test metrics have been well established for participants and follow-up research. In this paper, the case of resetting the vehicle service reminder system is used for illustration purposes. However, this new model can also be applied to other sub-systems that may not have a more user-friendly interface.

## 2. RELATED WORK

### 2.1. ODB and ODB II

The Electronic Control Module (ECM) is an embedded computer system that controls one or more of the electrical systems or subsystems in a motor vehicle [16]. Vehicles produced after the late 1980s are equipped with the On Board Diagnostic (ODB) system which is a computer controlled system for monitoring malfunctions. The ODB system uses a computer program to construct a malfunction table. Once a malfunction occurs, the system tracks and stores the malfunctioning information in ECM memory as code. There is a Malfunction Indicator Lamp (MIL) installed on the dashboard to indicate the appearance of a malfunction. This indicator tells a technician to use a reader to export one or more malfunction codes when it flashes [17].

This system is strengthened by ODB II for most model vehicles of 1994 and afterwards in the Europe and North America market [18]. The ODB II is able to monitor automobiles and diesel-engine vehicles, particularly for their emission systems. It standardizes the interface of the reader with an error code of five digits, which allows for the employing of computer-based approaches to find malfunctions, like using diagnostic fault tree to trace a malfunction [17]. This kind of On Board Diagnostic system makes it easy for a technician to output the malfunction code and identify the problem. Some reader devices in the On Board Diagnostic system also have basic input functionality, for example clearing the memory that stores the malfunction code(s). Therefore, an On Board Diagnostic system implements an interface between people, the vehicle for service, and maintenance purpose.

## 2.2. ECM tuning tool

ECM tuning tool is a flash reprogramming tool designed for both a vehicle performance tuner and engine management [19]. The original functionality of the ECM tuning tool is to manage the flow of fuel by tuning the spark control. The tool interacts with the ECM through the OBDII connector. It uses a USB interface to connect to a computer. The software running on the computer is able to retrieve vehicle programming information and can be re-programmed for other purposes, such as clearing Diagnostic Trouble Codes (DTCs) for all control modules in the vehicle [20]. Therefore, this system can also impact a service reminder with DTC support.

## 2.3. Keypad Keyless Entry System

For the Keypad Keyless Entry System, there is a keypad module that mounts in the door beneath the sheet metal, which can be seen as a small faceplate that is attached to the door face [21, 22]. This keypad is able to lock and unlock an automobile without using the vehicle key. To unlock (or lock) a vehicle, the driver must type in his (or her) personalized access code. Once a valid code is entered, the door will be unlocked (or locked). The use of this system has grown rapidly, which has resulted in derivative systems and devices such as the Remote Keyless Entry (REK) [23].

## 3. INPUT MODALITIES

The tools mentioned in the "Related Work" section, however, are not designed for every vehicle. For example, some vehicles do not possess a keypad keyless entry system. Another problem is that the operations of different devices within an automobile may vary from one vehicle to another, especially since every motor vehicle is not made by the same motor company. Therefore, it is important to have a universal system that can be used with any vehicle. A reasonable approach for acquiring such a universal system is to have a system that is all ECM oriented.

The motivation of i-Interaction is to design a driver-oriented model for driver-ECM interaction. The novel model utilizes direct input devices of a vehicle which are equipped on most vehicles. These devices could be used as input to the ECM and to let it serve in the manner that every automobile user feels comfortable when they reconfigure settings which are under the ECM's control. We find three different popular devices on a vehicle as i-Interaction input.

Most automobiles sold in the North American and European markets come with an in-dash radio receiver. This radio receiver plays radio, cassettes, CDs, and allows for in-car speakers by having an equalizer and amplifier. The control panel of a radio receiver usually comes with a power button, forward button, reverse button, and 0-9 memory buttons. These buttons are used to control the play of multimedia content by the drivers. When a driver pushes the button, the microcomputer of the radio receives a signal and changes the playing status.

Nowadays, advanced automobiles are manufactured with an in-dash navigation, information, and entertainment as one module of the central panel. This module uses a touch screen or a display screen with side buttons to show navigating information, vehicle running information, and multimedia contents. This allows the driver to either touch the screen or use available buttons to interact with the system. The input is sent to the embedded computer of the system which in turn responds to the user's input. When looking at the keypad keyless entry module, the i-Interaction model can employ its faceplate as the input device in automobiles that have this feature.

## 4. I-INTERACTION

Currently, a driver or a technician has to follow the reset procedure by manipulating the knob(s) on the instrument panel as the A and B in Figure 1 step by step. The following reset instruction is adapted from an automobile user manual.

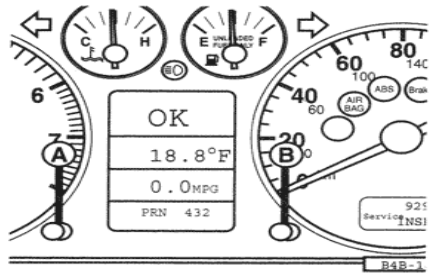

Figure 1. Knobs on Instrument panel of an Audi A4 (adapted from User's Manual of Audi)

Two example procedures for using the instrument panel to reset the system are:

Procedure A:

1) *Turn the ignition switch to the ON (II) position.*

2) *Push and release the Select/Reset knob repeatedly until the engine oil life indicator is displayed.*

3) *Press and hold the Select/Reset knob for about X1 seconds. The information display shows the reset mode display.*

4) *Press and hold the Select/Reset knob for another X2 seconds. The maintenance item code(s) will disappear, and the engine oil life will reset to 100.*

Procedure B:

1) *Switch ignition off.*

2) *Press trip odometer reset button (B) and at same time*

3) *Switch on the ignition. The central display window will show: "SERVICE IN XXXX" or "SERVICE".*

4) *Release button.*

5) *Turn clock adjuster knob (A) counter clockwise or clockwise couple of times to reset.*

In this circumstance, an operator first needs to read the chapter in the user manual which identifies the appropriate buttons/knobs on the instrument panel. Then, go to the service reminder chapter to follow the procedure. In order to follow the procedure step by step, the operator also needs to place the user manual in a position where he or she can read at the same time while manipulating the buttons/knobs. As an alternative to avoid this complicated process, some people prefer to take the negative battery cable off to clear the service reminder.

The i-Interaction model, on the other hand, borrows the input capability of the in-dash radio receiver or the in-dash all-in-one system as an input acquisition system for driver-ECM interaction such as the vehicle service reminder system. The i-Interaction model e program these numerical buttons as Figure 2 shows. So that, reference codes can be used to simplify the process of interacting with vehicle service reminder system. Using a reference code table to make entries takes the place of step-by-step instructions on how to reset a particular reminder in the user manual. Table 1 gives an example of the reference code table.

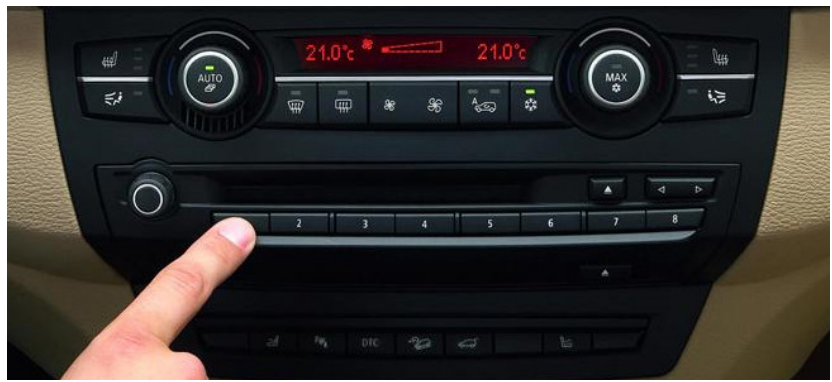

Figure 2. Numeric Button of an In-Dash Radio

Table 1. Reference Code Table

| Setting Item | Value | Code |
|---|---|---|
| Display Language | English | 1001 |
| | Spanish | 1002 |
| | French | 1003 |
| Time Zone | EST | 2001 |
| | CST | 2002 |
| | MST | 2003 |
| | PST | 2004 |
| Day Light Saving | Off | 2011 |
| | On | 2012 |
| Reset Service Reminder | Air Filter Clean | 3001 |
| | … … | |
| | Oil Change | 3014 |
| | Oil Filter & Oil Change | 3015 |

In the i-Interaction model, once a driver or a technician completes a maintenance work to a vehicle, what he or she needs to do is to look up the table, find the service item and its corresponding reference code from the manual. Next, turn the input button to be in the appropriate mode and key in the code. If the code is correct, the vehicle service reminder system will clear the reminder message of a required service which is posted by the vehicle service reminder system.

## 5. PROTOTYPE SIMULATION

The implementation of the i-Interaction simulation does not require expensive hardware or other complicated modifications. Basically, i-Interaction intercepts an input signal from the radio or the in-dash all-in-one system and is sent to the ECM to trigger the reset function of the service reminder system. For a usability test in our lab, we simulate the instrument panel as shown in Figure 2 where the keyboard only takes numeric key inputs as the numeric keypad of the in-car radio. The testing computer plays the role of ECM.

The simulation system renders an instrument panel on a computer monitor. A simulated LCD (on Bottom Left) indicates if a participant completes a reset operation. Automobile maintenance service includes a collection of post-service reminder reset operations. Instead of elaborating on each item in the entire i-Interaction list, we use the example of resetting an oil change reminder for illustrating the i-Interaction model. Engine oil is scheduled to change every 3000 miles or 3 months whichever comes first. When it is time to reset, the service indicator keeps reminding the driver with sound, light or text that flashes on the instrument LCD as indicated in Figure 3. A driver or a technician usually has to follow this procedure to reset the service reminder. In Figure 3, the simulated LCD shows that it is time to do an oil change for the automobile and this reminder needs to be reset after the oil change. In order to reset this reminder and other indicators, we print a user manual with reference codes for each. Figure 2, for example, shows this idea in the manner of automobile manual printing.

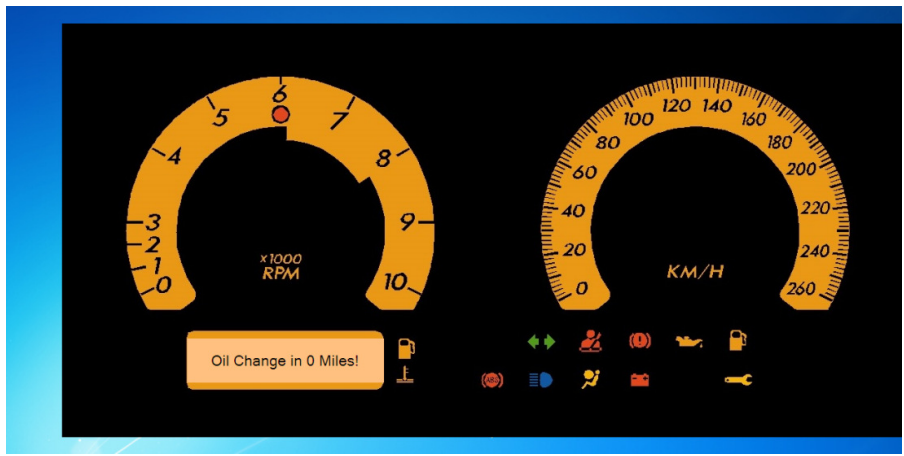

Figure 3. i-Interaction Simulator (service reminder)

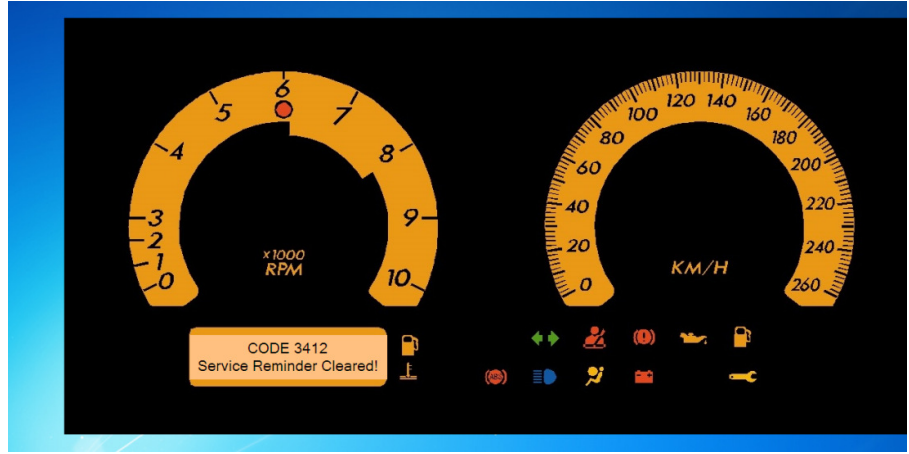

Figure 4. i-Interaction Simulator (reminder has been reset)

## 6. USABILITY TESTING (FUTURE WORK)

In order to test the usability of i-Interaction, we will ask subjects to undergo two procedures. The first procedure will ask for the participants to follow a similar conventional process for resetting the service reminder indicator as seen in Figure 1. The next procedure will then allow them to try the simulation system of the i-Interaction model to complete the same task. From these procedures, it is our goal to obtain feedback about the usability of i-Interaction as seen in Figures 3 and 4. There are three traits of usability that we want to measure from this test which are *time to complete, subjective satisfaction,* and *error rate by the users* [24]. The details for this procedure can be seen in the following subsection.

### 6.1. Detailed Description of the Future Study

In order to obtain participants for this study, there will be two qualifications that each person must meet. The first qualification is that the subjects must have a driver's license. For the second qualification, each participant must own or have access to an automobile on a daily basis. These requirements will not only ensure that the participants are familiar with operating an automobile but also eliminate any confounding factors that could affect the results from this study. As already mentioned, the selected participants will first complete the task of resetting the vehicle service reminder indicator by using a similar process to the procedure that involves using the knobs in the Instrument panel of an automobile. The second task will cause for them to undergo the same procedure but using the i-Interaction simulation.

There are four hypotheses (one main hypothesis; three sub-hypotheses) that will be either accepted or rejected in this study:

- **Main hypothesis:** *the i-Interaction model has a better usability than the Conventional model (using knobs from the instrument panel).*

- **Sub-hypothesis 1:** *i-Interaction is more efficient to use.*

- **Sub-hypothesis 2:** i-*Interaction is more satisfying to use.*

- **Sub-hypothesis 3:** *i-Interaction is less error prone.*

The main hypothesis will be answered through the acceptance or rejection of the three sub-hypotheses. If one of the sub-hypothesis is rejected, then the main hypothesis will be rejected.

Through the sub-hypotheses, there are three variables that will be measured respectively: *time to complete, subjective satisfaction,* and *error rate by the users.* To measure each variable, there will be different methods enacted. To measure the variable of *time to complete*, both simulations for each participant will be timed. There are two possible ways to conduct this procedure by either using time logging software [25] or a manual timer. The next variable, *subjective satisfaction*, will be measured through a survey. This survey will consist of questions that will be given to the participants to complete. It is the intent that the participants' feedback will determine the level of subjective satisfaction for both simulations. The final variable, *error rate by the users*, will be measured through our observance of the participants while using both simulations. We will observe and a keep track of the number of errors that have been made by the participants in both simulations. Once the study has been conducted and the sub-hypotheses are answered, the main hypothesis (the usability of the i-Interaction model) will then automatically be answered. Table 2 provides a summarized overview of the future study.

## 7. CONCLUSION

In this paper, we investigate the current in-vehicle interaction. We find that not all of the user-vehicle interaction is designed to be user-oriented. Therefore, we propose a new interaction model which allows direct user input. The i-Interaction model presents an intuitive interaction model for user with the In-Vehicle Information System (IVIS) by reprogramming the interface between user and IVIS. It is the belief that the i-Interaction approach will reduce effort and time to perform the operation in comparison to the traditional approach.

## ACKNOWLEDGEMENTS

Edward Dillon would like to acknowledge the U.S. Southern Regional Education Board for the funding opportunity to do research.

Table 2. Overview of the Future Study

| Summary of Usability Study | |
|---|---|
| **Simulations** | - Relative knobs on Instrument panels simulator<br><br>- i-Interaction simulator |
| **Participants** | The selection of participants will be based on two qualifications:<br><br>- Have a driver's license<br><br>- Own or have access to an automobile on a daily basis |
| **Hypotheses** | **Main Hypothesis:**<br><br>- the i-Interaction model has a better usability |
| | than the Conventional model (using knobs from the instrument panel) **Sub-Hypothesis 1:** - i-Interaction is more efficient to use **Sub-Hypothesis 2:** - i-Interaction is more satisfying to use **Sub-Hypothesis 3:** - i-Interaction is less error prone |
| **Variables being measured** | - Time to Complete - Subjective Satisfaction - Error Rate by User |
| **Procedures to measure variables** | **Time to Complete:** - use either time logging software or a manual timer to measure this task. **Subjective Satisfaction:** - a survey will be given to the participants that will measure their level of satisfaction with both simulations. **Error Rate by User:** - the facilitators of this study will observe and keep track of the number of errors made by the participants in both simulations. |
| **Goal of study** | Measuring the effectiveness of using the i-Interaction tool in for resetting the Vehicle Service Reminder System in automobiles when compared to using knobs on an Instrument panel. |

## REFERENCES

[1] Forlizzi, Jodi, Barley, William, & Seder, Thomas, (2010) "Where Should I Turn: Moving From Individual to Collaborative Navigation Strategies to Inform The Interaction Design of Future Navigation Systems", In *Proceedings of the 28th International Conference on Human Factors in Computing Systems* (Atlanta, Georgia, USA, April 10 - 15, 2010). CHI '10. ACM, New York, NY, 1261-1270.

[2] Iqbal, Shamsi, , Ju, Yun-Cheng, & Horvitz, Eric, (2010) "Cars, Calls, and Cognition: Investigating Driving and Divided Attention", In *Proceedings of the 28th International Conference on Human Factors in Computing Systems* (Atlanta, Georgia, USA, April 10 - 15, 2010). CHI '10. ACM, New York, NY, 1281-1290.

[3] Larry Constantine, Helmut Windl, (2009) "Safety, Speed, and Style: Interaction Design of an In-Vehicle User Interface", *In Proceedings of the 27th International Conference on Human Factors in Computing Systems* (Boston, MA, USA, April 4 – 9, 2009). CHI '09, ACM, New York, NY, 2675 – 2678.

[4] Bach, Kennet, Jæger, Mads, Skov, Mikael, & Thomassen, Nils, (2009) "Interacting with In-Vehicle Systems: Understanding, Measuring, and Evaluating Attention", In *Proceedings of the 2009 British Computer Society Conference on Human-Computer interaction* (Cambridge, United Kingdom, September 01 - 05, 2009). British Computer Society Conference on Human-Computer Interaction. British Computer Society, Swinton, UK, 453-462.

[5] Jensen, Brit, Skov, Mikael, & Thiruravichandran, Nissan, (2010) "Studying Driver Attention and Behaviour for Three Configurations of GPS Navigation in Real Traffic Driving", In *Proceedings of the 28th international Conference on Human Factors in Computing Systems* (Atlanta, Georgia, USA, April 10 - 15, 2010). CHI '10. ACM, New York, NY, 1271-1280.

[6] Froehlich, Peter, Schatz, Raimund, Leitner, Peter, Mantler, Stephan, and Baldauf, Matthias, (2010) "Evaluating Realistic Visualizations for Safety-Related In-Car Information Systems", In *Proceedings of the 28th of the international Conference Extended Abstracts on Human Factors in Computing Systems* (Atlanta, Georgia, USA, April 10 - 15, 2010). CHI EA '10. ACM, New York, NY, 3847-3852.

[7] Mancuso, Vince, (2009) "Take Me Home: Designing Safer In-Vehicle Navigation Devices", In *Proceedings of the 27th international Conference Extended Abstracts on Human Factors in Computing Systems* (Boston, MA, USA, April 04 - 09, 2009). CHI '09. ACM, New York, NY, 4591-4596.

[8] Forlines, Clifton, Schmidt-Nielsen, Bent, Raj, Bhiksha, Wittenburg, Kent, & Wolf, Peter, (2005) "A Comparison between Spoken Queries and Menu-based Interfaces for In-Car Digital Music Selection", *IFIP TC13 International Conference on Human-Computer Interaction (INTERACT)*, September 2005.

[9] Carter, Chris, & Graham, Robert, (2000) "Experimental Comparison of Manual and Voice Controls for the Operation of In-Vehicle Systems", *Proceedings of the IEA 2000/HFES 2000 Congress (CD-ROM)*, Santa Monica, CA: Human Factors and Ergonomics Society.

[10] Constantine, Larry, & Windl, Helmut, (2009) "Safety, Speed, and Style: Interaction Design of an In-Vehicle User Interface", In *Proceedings of the 27th international Conference Extended Abstracts on Human Factors in Computing Systems* (Boston, MA, USA, April 04 - 09, 2009). CHI '09. ACM, New York, NY, 2675-2678.

[11] Bach, Kenneth, Jæger, Mads, Skov, Mikael, and Thomassen, Nils, (2008) "You Can Touch, But You Can't Look: Interacting with In-Vehicle Systems", In *Proceeding of the Twenty-Sixth Annual SIGCHI Conference on Human Factors in Computing Systems* (Florence, Italy, April 05 - 10, 2008). CHI '08. ACM, New York, NY, 1139-1148.

[12] González, Ivan, Wobbrock, Jacob, Chau, Duen, Faulring, Andrew, & Myers, Brad, (2007) "Eyes on the Road, Hands on the Wheel: Thumb-Based Interaction Techniques for Input on Steering Wheels", *Proceedings of Graphics Interface 2007.*

[13]     Alpern, Micah, & Minardo, Katie, (2003) "Developing a Car Gesture Interface for Use as a Secondary Task", In *CHI '03 Extended Abstracts on Human Factors in Computing Systems* (Ft. Lauderdale, Florida, USA, April 05 - 10, 2003). CHI '03. ACM, New York, NY, 932-933.

[14]     Pirhonen, Antti, Brewster, Stephen, & Holguin, Christopher, (2002) "Gestural and Audio Metaphors as a Means of Control for Mobile Devices", In *Proceedings of the SIGCHI Conference on Human Factors in Computing Systems: Changing Our World, Changing Ourselves* (Minneapolis, Minnesota, USA, April 20 - 25, 2002). CHI '02. ACM, New York, NY, 291-298.

[15]     DeGraff, Brent L., (1998) "Navigation system with vehicle service information" US Patent No. 5,819,201.

[16]     Steel, Ryan, Haggitt, David, F. & Zielinski, J., (2008) "Electronic Control Module" US Patent Ref. 2008/0080147 A1.

[17]     Price, Chris, J., (1999) "Computer-Based Diagnostic Systems". Springer-Verlag, London, UK.

[18]     Bonnick, A. (2001) "Automotive Computer Controlled Systems", Butterworth-Heinemann, Oxford, MA, USA

[19]     "Turning Basics: What's it all about" http://tuningtools.co.uk/obd-tuning-information.html (accessed July 15, 2010).

[20]     OBD-II Trouble Codes.com: Your OBD-II Trouble Codes Repair Site, (2010) "OBD-II Trouble Codes Home" http://www.obd-codes.com/ (accessed July 15, 2010).

[21]     Donaldson, Edward, M. (1997) "Keyless entry system for replacement of existing key locks" US Patent No. 5,609,051.

[22]     Alrabady, Ansaf, I., & Mahmud, Syed, M (2003). "Some Attacks Against Vehicles' Passive Entry Security Systems and Their Solutions", *IEEE Transactions on Vehicular Technology*, Vol. 52, No. 2, March 2003, 431-439.

[23]     Lambropoulos, George, P. (1998) "Keyless Vehicle Entry and Engine Starting System" US Patent No. 5,736,935.

[24]     Shneiderman, Ben, (1998) "Designing the User Interface: Strategies for Effective Human-Computer-Interaction", Reading, Mass, Addison Wesley Longman.

[25]     Russinovich, Mark & Cogswell, Bryce, "Process Monitor v2.91", *Microsoft,* Published on May 19, 2010 (accessed July 16, 2010).